\documentclass[epj,twocolumn]{webofc}
\usepackage[varg]{txfonts}   
\usepackage{graphicx}

\wocname{epj}
\woctitle{Seismology of the Sun and the Distant Stars 2016}
\begin{document}
\title{A modulated RRd star observed by K2}
%

\author{\firstname{Emese} \lastname{Plachy}\inst{1}\fnsep\thanks{\email{eplachy@konkoly.hu}}, 
        \firstname{P\'eter} \lastname{Klagyivik}\inst{1},
        \firstname{L\'aszl\'o} \lastname{Moln\'ar}\inst{1} \and
        \firstname{R\'obert} \lastname{Szab\'o}\inst{1} 
}

\institute{Konkoly Observatory, MTA CSFK, Konkoly Thege Mikl\'os \'ut 15-17, H-1121 Budapest, Hungary}

\abstract{%
We report the analysis of the double-mode RR Lyrae star EPIC 205209951, the first modulated RRd star observed from space. The amplitude and phase modulation are present in both modes. 
}
\maketitle
%
\section{Introduction}
\label{intro}

Modulated (Blazhko) RRd stars have been recently discovered by the OGLE (Optical Gravitational Lensing Experiment) Survey in the Galactic Bulge \cite{ogle,smolec} and in the Magellanic Clouds \cite{anomal}. These stars belong to the "anomalous RRd variables" that have different period and amplitude ratios than typical double-mode RR Lyraes and the fundamental mode often dominates in their pulsation. The investigation of RR Lyrae stars in the globular cluster M3 also revealed Blazhko RRd stars with similar characteristics \cite{jurcsik}. Here we present EPIC 205209951, the first modulated RRd star observed in a space mission. This star was previously known as a strongly modulated RRab star: data from ground-based surveys revealed only the modulated fundamental mode that reaches higher amplitudes than the first overtone.


\section{Data and analysis}
\label{sec-1}

EPIC 205209951 was observed with 1-minute sampling mode in Campaign 2 of the K2 mission. We used the PYKE Software \cite{still} and the Extended Aperture Photometry (EAP, Plachy et al. these proceedings) to derive brightness data from the target pixel files. The main idea of the method is to determine the optimal  aperture that contains the star with the movements due to the attitude control manoeuvre corrections but avoid nearby stars. 

\begin{figure}
\centering
\includegraphics[width=\hsize,clip]{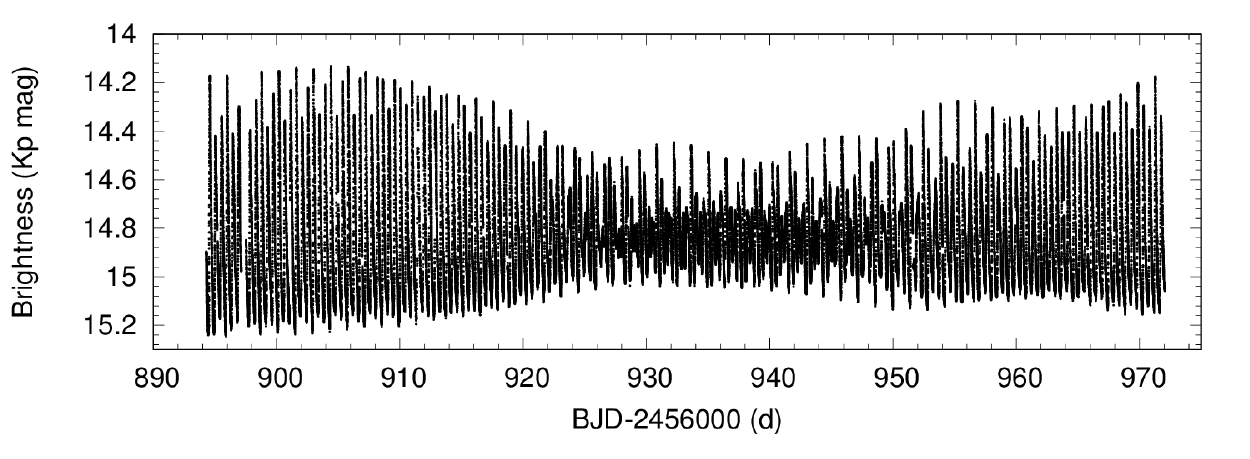}
\caption{The short cadence light curve of EPIC 205209951 obtained with EAP photometry.}
\label{fig-1}       
\end{figure}

\begin{figure}
\centering
\includegraphics[width=\hsize,clip]{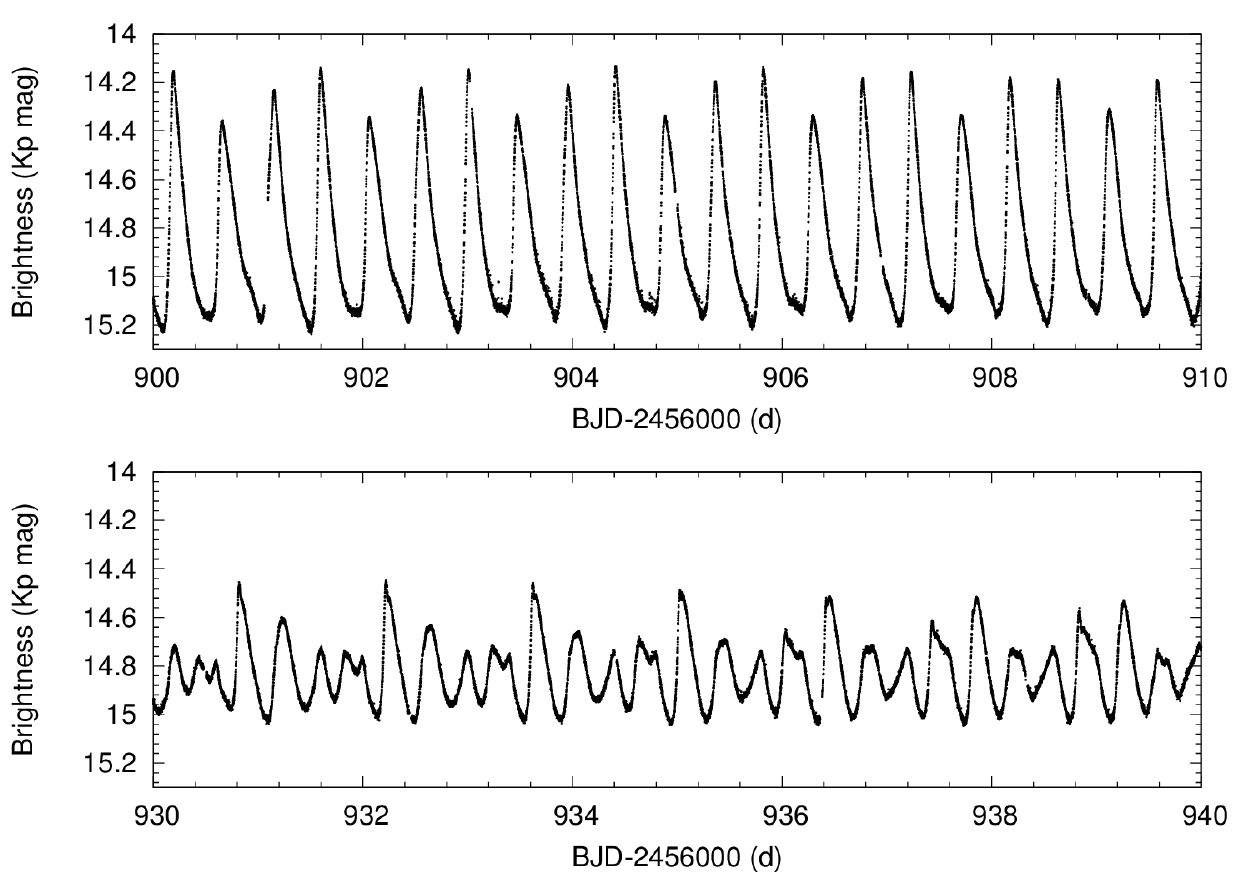}
\caption{Different sections of the light curve.}
\label{fig-2}       
\end{figure}

The EAP light curve of EPIC 205209951 is displayed in figure~\ref{fig-1}.
The figures shows how the changing amplitudes shape the light curve. The middle part resembles a classical RRd star, but elsewhere it is dominated by the fundamental mode and looks very different (see figure~\ref{fig-2}).

Period04 software \cite{period04} was used to calculate the Fourier transform and the time variation of the pulsation modes. The fundamental mode and the first overtone  frequencies were determined to be  $f_0$=2.124311  c/d and $f_1$=2.86714 c/d. The fundamental mode has a higher average amplitude than the first overtone. While in most RRd stars the amplitude ratio is the opposite, this is typical for the anomalous RRd stars. The $\sim$0.741 period ratio is also slightly outside the canonical 0.742–0.748 range for the double-mode RR Lyrae period ratios. 

\begin{figure}
\centering
\includegraphics[width=\hsize,clip]{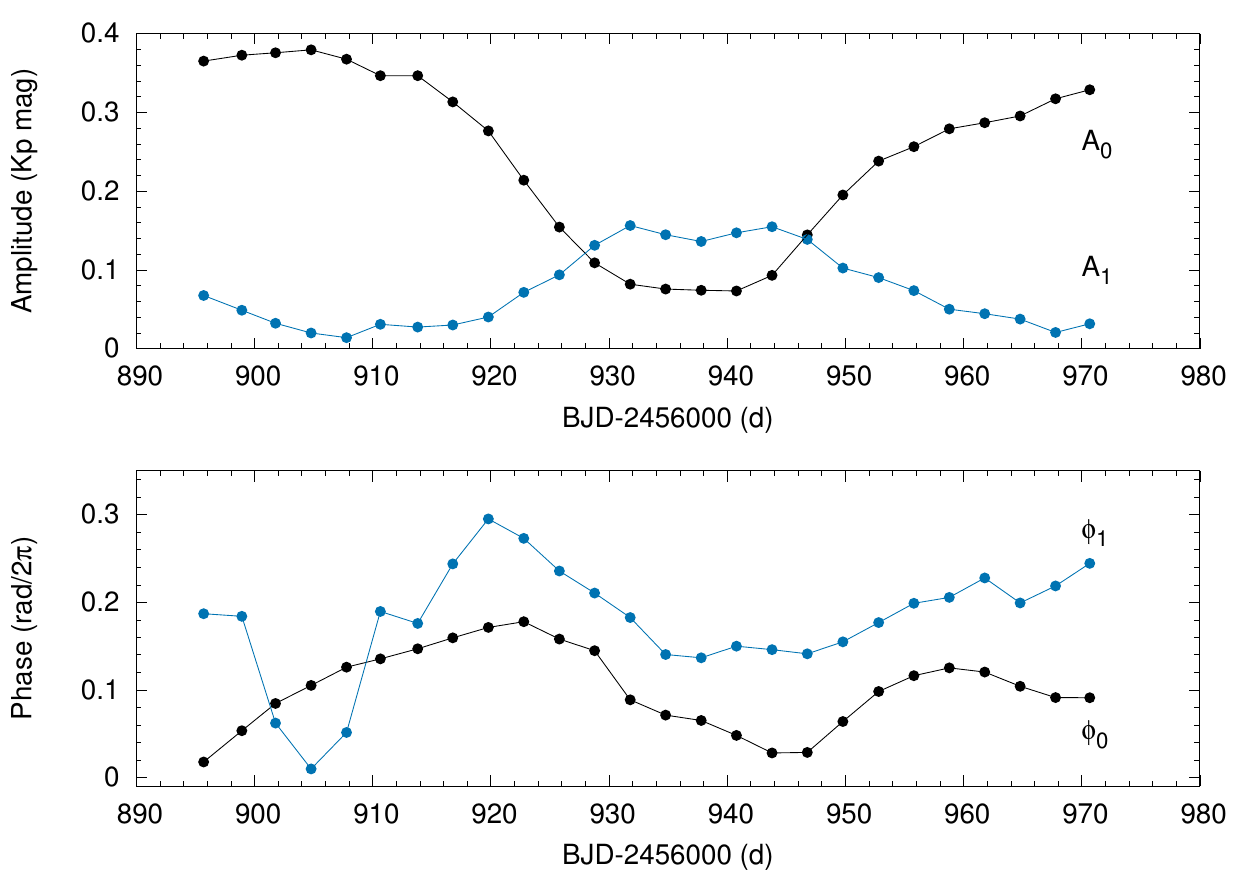}
\caption{The temporal variation of the amplitudes and phases of the different the fundamental mode (black) and the first-overtone mode (blue) calculated with the Period04 software using 3 day bins.}
\label{fig-3}       
\end{figure}

The temporal variations of the $A_0$, $A_1$, $\phi_0$ and $\phi_1$ parameters indicate that both modes are modulated, and likely not independently. While the modulation amplitudes show clear anticorrelation, the phases are nearly correlated except in the first twenty days of the observations, as it is seen in figure~\ref{fig-3}.

\section{Conclusions}

EPIC 205209951 has all the characteristics that anomalous RRd stars do:
the unusual period ratio, the dominance of the fundamental mode in the amplitude and the modulation. Thanks to the high-precision data provided by \textit{Kepler}, a new member of this strange group can be examined in great detail. The preparation of a detailed investigation of the light curve characteristics and frequency content of  EPIC 205209951 is in progress (Plachy et al. in prep).

\label{sec-con}
\section*{Acknowledgements}
\label{sec-ack}

This work has used K2 targets selected and proposed by the RR Lyrae and Cepheid Working Group of the Kepler Asteroseismic Science Consortium (proposal number GO4069). Funding for the Kepler and K2 missions is provided by the NASA Science Mission directorate. This project has been supported by the LP2014-17 Program of the Hungarian Academy of Sciences, and by the NKFIH K-115709, PD-116175 and PD-121203 grants of the Hungarian National Research, Development and Innovation Office. L.M. was supported by the János Bolyai Research Scholarship of the Hungarian Academy of Sciences. The research leading to these results has received funding from the European Community's Seventh Framework Programme (FP7/2007-2013) under grant agreement no. 312844 (SPACEINN). The authors also acknowledge support from to ESA PECS Contract No. 4000110889/14/NL/NDe. This work made use of PyKE \cite{still}, a software package for the reduction and analysis of Kepler data.


\begin{thebibliography}{}

\bibitem{period04}
Lenz, P., Breger, M.,  Communications in Asteroseismology \textbf{146}, 53 (2005)
\bibitem{still}
Still M., Barclay T., Astrophysics Source Code Library, record
ascl:1208.004 (2012)
\bibitem{anomal}
Soszy{\'n}ski I. et al., MNRAS \textbf{463}, 1332 (2016)
\bibitem{smolec}
Smolec, R. et al., MNRAS \textbf{447}, 3756(2015)
\bibitem{ogle}
Soszy{\'n}ski I. et al., AcA \textbf{64}, 177 (2014)
\bibitem{jurcsik} 
Jurcsik, J. et al., ApJ \textbf{797}, 3 (2014)
\end{thebibliography}
%
%

\end{document}